Table of Contents Graphic

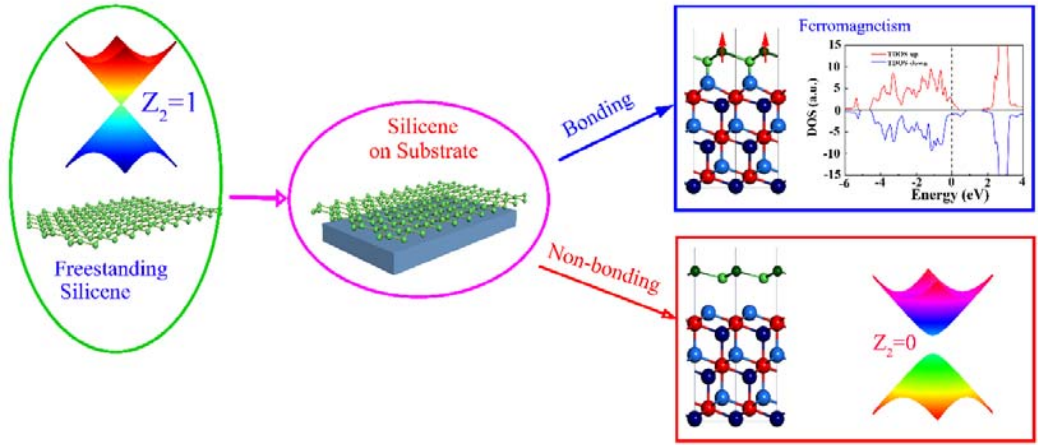

# Substrate-induced magnetism and topological phase transition in silicene

Ke Yang [1], Wei-Qing Huang [1, *], Wangyu Hu [2], Gui-Fang Huang [1, #], Shuangchun Wen [1]

[1] *Department of Applied Physics, School of Physics and Electronics, Hunan University, Changsha 410082, China*

[2] *School of Materials Science and Engineering, Hunan University, Changsha 410082, China*

**Abstract**: Silicene has shown great application potential as a versatile material for nanoelectronics, particularly promising as building block for spintronic applications. Unfortunately, despite its intriguing properties, such as relatively large spin-orbit interactions, one of the biggest hurdles for silicene to be useful as a host spintronic material is the lack of magnetism or the topological phase transition owing to the silicene–substrate interactions, which influence its fundamental properties and has yet to be fully explored. Here, we show that when silicene is grown on $CeO_2$ substrate, an appreciable robust magnetic moment appears in silicene covalently bonded to $CeO_2$ (111), while a topological phase transition to a band insulator occurs regardless of van der Waals (vdWs) interaction or covalent bonding interaction at interface. The induced magnetism of silicene is due to the breaking of Si-Si π-bonding, also resulting in trivial topological phase. The silicene-substrate interaction, even weak vdWs force (equivalent to an electric field), can destroy quantum spin Hall effect (QSHE) of silicene. We propose a viable strategy——constructing inverse symmetrical sandwich structure (protective layer/silicene/substrate)——to preserve quantum spin Hall (QSH) state of silicene in weak vdWs interaction system. This work takes a critical step towards fundamental physics and realistic applications of silicene-based spintronic devices.

[*]. Corresponding author. *E-mail address:* wqhuang@hnu.edu.cn
[#]. Corresponding author. *E-mail address:* gfhuang@hnu.edu.cn

# I. Introduction

Two-dimensional (2D) materials (such as graphene and silicene) with a honeycomb geometry have attracted increasing attention owing to their unique electronic properties [1-6]. Ever since the prediction of quantum spin Hall (QSH) state in freestanding silicon monolayer (silicene), researchers have been infatuated with the idea of its applications in spintronic device using its topologically protected dissipationless edge states, because of its uniquely suitable for integration in Si-based electronics. Compared to graphene, silicene possesses a relative large intrinsic spin-orbit coupling (SOC), opening a band gap (~1.6 meV [7]) at Dirac points and topological nontrivial electronic structure, which can host the quantum spin Hall effect (QSHE) up to 18 K [7], suggesting it is a good candidate for spintronic device.

However, in sharp contrast to graphene, which can be mechanically exfoliated from graphite, silicene has to be grown on a substrate, because the $sp^3$ hybridization of Si atoms is more favorable than the $sp^2$ hybridization [8]. This results into a formidable obstacle to seeking appropriate substrate to stabilize and simultaneously preserve the outstanding properties of silicene. To date, only certain corrugated silicene phases has been experimentally grown on some substrates, such as Ag (111 and 110) [9-14], Ir (111) [15], Ru (0001) [16], $ZrB_2$ (0001) [8] and ZrC (111) [17]. However, the silicene-substrate interaction is so strong that it does not only markedly change the silicene structure, but also destroy intrinsic electronic nature of silicene [18]. For instance, the orbital hybridization between Ag and Si atoms leads to a surface metallic band and depresses the Dirac fermion characteristics [14] in the epitaxial silicene on Ag(111) surface [19] [20]. In principle, eliminating or minimizing substrate effects could preserve Dirac cone in silicene [21-24]. Until now, only a few substrates (for instance, $MgX_2$ (X=Cl, Br and I) [25], GaS [26], $CaF_2$ [23], and BN [27, 28]) can approximately preserve Dirac cone of silicene. An effective strategy to weaken the strong silicene-substrate interaction is by intercalation, as illustrated by quasi-freestanding silicene obtained via oxygen intercalation into bilayer silicene on Ag (111), which restores the Dirac cones [29]. Unfortunately, the band gap of silicene on the substrates is significantly enlarged [30, 26, 29, 27, 23, 25] compared to freestanding silicene, even although the silicene-substrate interaction is very weak. The large band gap induced by substrate indicates that silicene has occurred topological phase from topological insulator to band insulator [27, 31], leaving its topologically protected dissipationless edge states out of scope of the possible spintronic applications.

The lack of intrinsic magnetism is another hurdle for silicene to be useful as a spintronic material because of the absence of *d* or *f* electrons. A prerequisite for facilitating its potential applications in nanosized spintronic devices is to induce magnetism in silicene. It is shown theoretically that atomic scale defects in silicene phases, e.g., adatoms and vacancies, can carry magnetic moments μ of about several Bohr magneton, $\mu_B$ [32-35]. Moreover, extended defects such as zigzag edges can give rise to antiferromagnetic magnetism [4, 36, 37]. The most direct approach to realize magnetic properties in silicene is intercalation (or adsorption) by inherently magnetic metal atoms with strong magnetic properties coming from the unfilled shell electrons[33]. All these leave little doubt that magnetism in silicene phases can in principle to be induced, although the whole subject remains under active investigation, especially as concerns (i) whether intrinsic magnetism in silicene can be induced or not and (ii) the effect of silicene-substrate interaction on its magnetism as it is hard to produce a freestanding silicene with high stability and reliability.

The recent interest in silicene/substrate system has inevitably led to the question of possible their interactions and their underlying mechanism, especially due to the fact that the interfacial interactions dominate the fundamental properties of silicene. Generally, the silicene-substrate interaction can be classified into two types: van der Waals (vdW) interaction and covalent bonding interaction. The theoretical studies of silicene/substrate system are so far focused on their weak van der Waals (vdW) interactions [30, 21, 26, 38, 22-25], except one exception that a predicted corrugated silicene phase with significant structural reconstruction on SiC (0001) includes isolated Si−C pairs at interface [39]. The covalent bond at interface between the corrugated silicene phases and substrates (for example, Ag (111)) has been experimentally demonstrated [18]. Obviously, the research to explore an appropriate substrate to grown 'real' (uncorrugated) silicene and understand the silicene-substrate interaction is still in its infancy with investigations only on a few isolated substrates. Particularly, the latter is the prerequisite and foundation for the former, which has to be selected using exhaustive trial-and-error studies. Therefore, the understanding of the nature of the silicene-substrate interaction and its effect is of the utmost interest.

Herein, we first propose that $CeO_2$ is an ideal substrate, which offer a platform to systematically investigate the strong covalent and weak vdWs interactions between silicene and substrate. Firstly, $CeO_2$, as an oxide with wide band gap, has appropriate work function, enabling its band edge far from the Dirac-cone state in silicene. Secondly, the lattice mismatch between $CeO_2$ (111) and silicene is very small, avoiding local unequivalence Si atoms stress in their hybrids; thus the Rashba SOC is not too strong to annihilate the QSHE in silicene[40]. This is in sharp contrast to other substrates, such as Ag (111) [10, 12, 13] and SiC (0001) [39, 41], which lead to the structural reconstruction of silicene due to their large lattice mismatch. Thirdly, being a high dielectric oxide protective layer in silicon-based device, $CeO_2$ has usually been epitaxially grown on in Si (111) surface in mature semiconductor technology [42-45], indicating that the epitaxial growth of silicene on $CeO_2$ surface is possible. Large-scale *ab initio* calculations reveal that for silicene-$CeO_2$ system, either covalent or dispersive (vdWs) interactions at interface can exist, depending on their relative orientations. We show that silicene covalently bonded to $CeO_2$ (111) surface shows intrinsic magnetism, due to the fact that the chemical bond at silicene-substrate interface disrupts the π-bonds of silicene. The weak vdWs interaction between silicene and $CeO_2$, equivalent to an external electric field, can annihilate the spin-orbit coupling effect in silicene, leading to the silicene becoming a band insulator. Unfortunately, external electric field seems impossible to recover the non-trivial topological states of silicene due to the screen effect of substrate. We here propose a viable strategy――constructing inverse symmetrical sandwich structure (protective layer/silicene/substrate)――to preserve quantum spin Hall (QSH) state of silicene, which is demonstrated by using two representatives ($CeO_2$ (111)/silicene/$CeO_2$ (111) and $CaF_2$ (111)/silicene/$CaF_2$ (111)) through the calculated edge states and $Z_2$ invariant. This work takes a critical step to reveal the interfacial interaction between silicene and substrate for fundamental physics and realistic applications of silicene-based nanoelectronic devices.

## II. Computational Methods

Density functional theory (DFT) was performed to achieve optimized geometrical and electronic structures with a projector augmented wave (PAW) [46] basis as implemented in Vienna Ab Initio Simulation Package code [47, 48]. The Perdew-Burke-Ernzerhof (PBE) Generalized Gradient Approximation (GGA) exchange-correlation functional method was adopted. All calculations were performed using the DFT/GGA+U method (Ce 4*f* and O 2*p* were 9.0 and 4.5 eV) to obtain band gap of $CeO_2$. Moreover, the spin orbit coupling (SOC) was taken into account and the van der Waals interaction (DFT-D3 method with Becke-Jonson damping) was incorporated [49, 50]. The kinetic energy cutoff was 500 eV. Brillouin zone integration was performed on grids of 15 × 15 × 1 Monkhorst–Pack k-points. Total energy and all forces on atoms converged to less than $10^{-8}$ eV and 0.005 eV/Å. The vacuum space of 20 Å along the z direction is used to decouple possible periodic interactions. The binding energy also been estimated by $E_b = E_{tot} - E_{silicene} - E_{substrate}/N$, where, $E_{tot}$, $E_{silicene}$ and $E_{substrate}$ are the total energy of silicene/substrate, silicene and substrate, respectively, $N$ is the numbers of Si atoms in silicene.

## III. RESULTS AND DISCUSSION

**Dispersive and covalent interactions between silicene and $CeO_2$ (111).** As silicene is put on $CeO_2$ (111) surface, four different stacking patterns with high symmetry are obtained, as shown in Fig. 1. By closely inspecting the optimized structures, two different kinds of interactions (i.e., covalent and noncovalent interactions) are unexpectedly appeared at the silicene-$CeO_2$ (111) interface, which are different from those in silicene and other substrates [21, 26, 23, 51]. As the lower layer Si atoms in silicene are positioned on top of the topmost O atoms in $CeO_2$ (111) surface, the covalent bond between Si and O atoms is formed. The covalently bonded silicene/$CeO_2$ (111) system can be subdivided into CA configuration (the upper layer Si atoms are located on top of the second layer O atoms, Fig. 1 (a1)) and CB configuration (the upper layer Si atoms are positioned on top of the Ce atoms, Fig. 1 (a2)). For the noncovalent silicene/$CeO_2$ (111) system, the upper (lower) layer Si atoms positioned on top of Ce atoms is denoted as NA (NB) configuration (Figs. 1(b1) and (b2)). The equilibrium distance between silicene and $CeO_2$ (111) also reflects different interfacial interaction types: the equilibrium distance (~1.76 Å) at the bonded interface is much shorter than those (2.57~2.83 Å, Table I) at the non-bonded interface.

The former is the Si-O bond length at interface, while the latter corresponds to vdWs rather than covalent distance.

To assess the stability of these structures, the binding energies of NA and NB configurations are calculated to be -374.3 and -377.7 meV/Si respectively, indicating that they have the almost same thermodynamic stability at room temperature and possibility to be synthesized. For the covalently bonded silicene/$CeO_2$ (111) system, the binding energy is more negative (-1.23 and -1.38 eV/Si for CA and CB configuration, respectively), implying that silicene is much easier to grow on $CeO_2$(111) surface than on Ag(111) surface (< 0.9 eV/Si, [52]), which has been widely confirmed in experiments. Furthermore, *ab initio* molecular dynamics (AIMD) simulations have been performed at room temperature (300 K) to demonstrate whether silicene on $CeO_2$ (111) surface are indeed thermodynamically stable or not. A canonical ensemble was adopted for the AIMD simulations, with the time step of 2 fs. A 3×3×1 supercell containing 99 atoms for each configuration was adopted, and the structures of four configurations after annealing 5 ps are displayed in Fig. 2 and S1 (a1 and b1). Clearly, the buckled shape of silicene on the $CeO_2$ (111) surface, regardless of their interactions, is more stable than that of freestanding silicene (Fig. S2) after 5 ps AIMD simulations, suggesting the $CeO_2$ substrate can enhance the structural stability of silicene. Particularly, the Si atoms in the covalently bonded silicene/$CeO_2$ (111) system (CA and CB configurations) are only slightly distorted, because of the strong covalent interaction between Si-2 and O-1 atoms at interface.

To discuss the dynamic stability of the silicene/$CeO_2$ (111) system, we study their lattice dynamics by calculating their phonon dispersion by Phonopy code [53]. The results are presented in Figs. 2 and S1. The absence of imaginary modes in the entire Brillouin zone confirms that the silicene/$CeO_2$ (111) system is dynamically stable, which is in agreement with the results of AIMD. The phonon spectra of vdWs silicene/$CeO_2$ (111) system are almost identical, and is also for the covalently bonded ones; whereas the discrepancy between them further shows their different interfacial interactions. A remarkable phonon gap can be observed in the phonon spectra of the covalently bonded silicene/$CeO_2$ (111) system, in which a high frequency mode (about 19 THz) appears (Fig. 2 (a2) for CA and Fig. S1 (a2) for CB). Detailed analysis of the atom-resolved phonon density of states (PhDOS) reveals that the Si-2 -- O-1 bond at interface is predominant in the dispersionless high-frequency modes (Figs. 2 (a3) and S1 (a3)), showing the characteristics of *sp³* hybrid of Si atoms. More precisely, the high frequency phonon mode mainly derives from the vibration of Si-2 and O-1 atoms only along the direction of their bond (Figs. 2 (a3, a4) and S1 (a3, a4)): their vibration intensity along z direction is drastically strong compare to other Si, O and Ce atoms. This is due to the short Si-2-O-1 bond length of 1.76 Å, much smaller than that of O-Ce (2.43 Å) or Si-Si (2.35 Å) bond length. The high and distinguishable frequency vibration mode will be reflected in the Raman or infrared spectrum, implying that the interaction type in the silicene/$CeO_2$ (111) system can be easily identified by Raman or infrared spectrum measurement as silicene is grown on the $CeO_2$ (111) surface.

A first understanding of the bond mechanism between silicene and $CeO_2$ (111) can be collected from the structural symmetry and electronic structure. In bulk $CeO_2$, the Ce and O atoms are eight- and four-fold coordinated, respectively; and the O atoms locate at the center of the regular tetrahedron ligand field formed by four nearest neighbor Ce atoms. As a result, the O atoms can form *sp³* hybridization and bond with four nearest neighbor Ce atoms. However, the O atoms on the clean $CeO_2$ (111) surface are lost a ligand Ce atom, thus its $p_z$ electron is unsaturated. The unsaturated O-1 atoms prefer to bond with to the Si atoms when they approach each other in CA and CB configurations. Figs. 3 (a) and S3 (a) illustrate that there exists a Si-2–O-1 hybridization, as can be recognized from the occurrence of the partial densities of state (PDOS) peaks at the same energies for Si-2 $3p_z$ and O-1 $3p_z$ states, indicating that the σ-bonding is formed at interface. The bonding characteristics can be visualized in a very intuitive way, by the electron localization function (ELF), which is shown in Figs. 3 (c-d) and S3 (c-d). Here, ELF values vary from of 0.5 for free electrons to 1 for fully localized electrons, and those between 0.7-0.8 indicate a covalent bond character. Figs. 3 (c) and S3 (c) show that the electrons are distributed between Si-2 and O-1 atoms, verifying the O-1-Si-2 bond is formed. Therefore, the π-bonding network in silicene covalently bonded to $CeO_2$ (111) is broken, leaving $3p_z$ electrons of Si-1 atoms unpaired and localized. This is distinctly different from the vdWs silicene/$CeO_2$ (111) system, in which the π-bonding network in silicene is clear (Figs. 3 (d) and S3 (d)). In the covalent silicene/$CeO_2$ (111) system, the energy level of unsaturated $p_z$ electron of the Si-1 atom is higher than those of other Si atoms (Fig. 3 (a)). This leads to the $3p_z$ orbital of the Si-1 atom cross the Fermi level in the band structure (Fig. 3 (b)), giving rise to the metallic behavior in the covalently bonded silicene/CeO$_2$ (111) system (CA and CB configurations). Three-dimensional charge density differences (Fig. S4) also reflect the strong covalent bonding and weak vdWs interaction characteristics in silicene/CeO$_2$ (111) system. Bader charge analysis reveals that the Si-2 atom loses about one electron and the O-1 atom gains more than one electron compared to other O atoms. Interestingly, the different interactions at interface will lead to novel and distinct properties, such as magnetism in silicene, which will be discussed next in detail.

**Substrate-induced magnetism in silicene covalently bonded to CeO$_2$ (111).** The stable structure of isolated silicene has a low-buckled configuration because of the tendency of silicon atoms to adopt *sp$^3$* and *sp$^2$* hybridization over only *sp$^2$* hybridization and its weakened π bonding of the electrons in the outer shell. The low-energy electronic properties of silicene are determined by the outer 3*p$_z$* orbitals forming a weakly and extensive π-bonding network. The resulting delocalized π electrons result into a metallic and nonmagnetic (NM) monolayer silicene. When half of the Si atoms (lower-layer Si atoms) are bonded to the top O atoms in CeO$_2$ (111), strong σ bonds are formed between Si-2 and O-1 atoms and the π-bonding network is broken, leaving the electrons in the unsaturated Si-1 atoms localized and unpaired, and thus resulting in Si-1 being spin polarized with an appreciable magnetic moment. The spin-resolved total DOSs for the covalently bonded silicene/CeO2 (111) system (CA and CB configurations) are shown in Figs. 4 and S5 (a), in which it is clear that the magnetism occurs in these systems because there is an asymmetry between spin-up and spin-down DOSs. This is the first to reveal the function of substrate to induce the magnetism in silicene while an appropriate interface interaction is achieved in silicene/substrate system. The total magnetic moment in CA configuration is calculated to be about 1.2 μ$_B$, indicating that it is not merely derived from one unpaired electron on Si-1 atom, which carries only 1 μ$_B$. To explore the origin of magnetism, the spin-resolved partial DOSs of the covalently bonded silicene/CeO$_2$ (111) system are presented in Figs. 4(b) and S5 (b). One can see that the states near the Fermi level are mainly contributed by the *p* electrons of Si-1 atom, mixing with some *p* electrons of those atoms at interface (O-1 and Si-2 atoms). The large transfer of charge at interface causes the O-1 atom departing the center position of tetrahedron ligand field formed by three nearest neighbor Ce atoms and Si-2 atom, giving rise to the extra magnetic moment. To illustrate the space distribution of spins, the 3D electronic spin density ($\rho_{\uparrow(r)} - \rho_{\downarrow(r)}$) are plotted in Fig. 4 (c). Obviously, the substrate-induced magnetism in silicene is mainly localized around the unsaturated Si-1 atom, in agreement with the charge above Fermi level mainly from $p_z$ orbit of Si-1 (Fig. 4 (d) and (e)). We have also calculated a (2×2×1) supercell with SOC, the similar results are obtained. This can be understood by analyzing the character of $p$ electrons. Because the $p$ electrons are more delocalized than those in $d$ or $f$ states to some extent, there exists a long-range ex-change coupling between the $p$ electrons, which is responsible for the induced long-range ferromagnetic coupling in the system [34, 54, 35, 55]. Compared to other methods, such as part functionalized silicene and element doping [34, 54, 35, 37, 55], utilizing substrate to induce magnetism in silicene is a more feasible and cost-effective strategy, which is critical for the applications of silicene in nanoscale devices.

**Topological phase transition in silicene physically adsorbed on $CeO_2$ (111)**. Understanding the vdWs interaction between silicene and substrate is a most important task since it may have drastic consequences for the electronic structure and topological properties of silicene. As discussed above, a substrate-induced topological phase transition from topological insulator to band insulator by the Si-O bond at interface has been demonstrated by the orbitals of Si atom cross the Fermi level and the disappearance of Dirac cone in silicene covalently bonded to $CeO_2$ (111) (Fig. 3(b) and S3 (b)). In contrast, it is assumed that the intrinsic electronic properties of silicene could be preserved when silicene is affected by the weak vdWs interaction [30, 24]. Figs. 5(a) and S6 (a) show that the approximate linear Dirac cone in silicene is still clear in NA and NB configurations. Moreover, its Dirac cone states near fermi level are far from (about 1.5 eV in energy) the valence band maximum and conduction band minimum of $CeO_2$, demonstrating that $CeO_2$ is an ideal substrate for growth of silicene [30]. Meanwhile, a band gap of silicene is opened at Dirac cone (245.2 and 286.1 meV for NA and NB configurations, respectively, Table I), like the cases that silicene is grown on other substrates [30, 38, 24].

To reveal the physical origin of band gap opening in silicene, we have constructed the low-energy effective model of silicene on substrates presenting of SOC near the K points by tight-binding methods [40, 56]:

$$H = -t\sum_{<i,j>\alpha} c_{i\alpha}^\dagger c_{j\alpha} + i\gamma_{so}\sum_{\ll i,j\gg\alpha\beta} v_{ij} c_{i\alpha}^\dagger \sigma_{\alpha\beta}^z c_{j\beta} + i\frac{\gamma_{R1}(E_z)}{2}\sum_{<i,j>\alpha\beta} c_{i\alpha}^\dagger (s\times \hat{d}_{ij})_{\alpha\beta}^z c_{j\beta} -$$

$$i\gamma_{R2}\sum_{\ll i,j\gg\alpha\beta} \mu_{ij} c_{i\alpha}^\dagger (s\times \hat{d}_{ij})_{\alpha\beta}^z c_{j\beta} + i\gamma_m \sum_{<i,j>\alpha\beta} \mu_i c_{i\alpha}^\dagger c_{j\alpha} + \gamma_B \sum_{i\alpha\beta} c_{i\alpha}^\dagger \sigma_{\alpha\beta}^z c_{j\beta} \quad (1)$$

where $c_{i\alpha}^\dagger$ ($c_{j\beta}$) creates $c_{j\beta}$ an electron with spin polarization at $\alpha$ site $i$, and $<i,j>/\ll i,j\gg$ run over all the nearest or next-nearest neighbor hopping sites. The first term represents the usual nearest-neighbor hopping. The second term represents the effective SOC that contains the intrinsic "atomic" SOC term of monolayer silicene plus $\gamma_{so}^{ind}$ which can been induced by the substrate, where $s = (s_x, s_y, s_z)$ is the Pauli matrix of spin, with $v_{ij} = \pm 1$ is clockwise or anticlockwise of next-nearest-neighboring hopping with respect to the positive z axis. The third term represents the first Rashba SOC, which is induced by external electric field or substrates, and $\hat{d}_{ij} = d_{ij}/|d_{ij}|$ with the vector $d_{ij}$ connecting two sites $i$ and $j$ in the same sublattice. The forth term represents the second Rashba SOC associated with the next-nearest neighbor hopping term, where $\mu_{ij} = \pm 1$ for the A (B) site, this term is negligible for silicene, so we set $\gamma_{R2} = 0$ in this paper. The fifth term is the staggered sublattice potential term, which describes the breaking of the sublattice symmetry by the interaction with the substrates. The sixth term represents the "pseudomagnetic" term. We expand the TB Hamiltonian surrounding the two valley K points, and obtain the low-energy effective model:

$$H_k = \hbar v_F(\sigma_x k_x + \sigma_y k_y) + \gamma_m s_z + \frac{\gamma_{R1}}{2}(\sigma \times s) + \gamma_{so}\sigma_z s_z + \gamma_B s_z \quad (2)$$

where $v_F = \frac{\sqrt{3}t}{2}$, it are listed Table I. $\sigma$ is the Pauli matrices of AB- sublattice.

From the Eq. (2), the parameters are:

$$\gamma_m = \frac{(\varepsilon_4 - \varepsilon_3) + s(\varepsilon_2 - \varepsilon_1)}{4} \quad (3.1)$$

$$\gamma_{R1} = \pm\frac{(\varepsilon_2 - \varepsilon_1)\sqrt{1-s^2}}{2} \quad (3.2)$$

$$\gamma_{so} = \frac{(\varepsilon_4 + \varepsilon_3) - (\varepsilon_2 + \varepsilon_1)}{4} \quad (3.3)$$

$$\gamma_B = \frac{(\varepsilon_4 - \varepsilon_3) - s(\varepsilon_2 - \varepsilon_1)}{4} \quad (3.4)$$

where $s$ is the expectation values for the z component of spin, which can been acquired by the wavefunction from DFT [40]. These parameters can easily derived by the calculated energy at K point using the DFT, which are listed in Table I.

The band structure near the Dirac point from the low-energy effect Hamiltonian is quite close to the results of DFT (Figs. 5 (b) and S6 (b)), indicating that the tight-binding model can describe the electronic properties of silicene on substrates well. Clearly, the band gap of silicene on $CeO_2$ is increased compared with that of freestanding silicene, which can mainly be attributed to the "mass" terms ($\gamma_m$) appearing. The large "mass" terms (119.6 and 142.1 meV for NA and NB configurations, respectively; Table I) imply that the substrate effect is equivalent to a perpendicular electric field, which its magnitude depends on stacking patterns. Direct DFT calculation shows that the static electronic potential difference between upper and lower layer Si atoms are 231.1 and 295.4 meV for NA and NB configurations, in well agreement with the sublattice potential step ($2\gamma_m$) by tight-binding model. Unexpectedly, the spin-orbit term is changed due to substrate effect: its magnitude relies on the buckle height of silicene and stacking patterns; in particular, the bigger the buckle height (or sublattice), the stronger the spin-orbit interaction will be [40, 31]. Compared with "mass" terms, however, the change of spin-orbit term is slight. To compare the intrinsic "atomic" SOC term of monolayer silicene and the SOC induced ($\gamma$) by $CeO_2(111)$ surface, we construct ideal silicene/$CeO_2(111)$ models, which are only varied the interfacial distance and without optimizing structure of silicene (named as NA-H or NB-H configuration). The calculated $\gamma_{so}^{ind}(=\gamma_{so}^{h-NA}-\gamma_{so}^{silicene})$ are 0.67 and 0.20 $meV$ for NA and NB configurations, respectively, indicating the substrate increases the intrinsic SOC induced gap of silicene. Due to breaking of reflection symmetry, $CeO_2$ (111) surface also induces the extrinsic Rashba SOC, which its strength (Table I) is positive correlation with the sublattice potential and intrinsic SOC, in agreement with others [31]. A "pseudomagnetic" term, induced by $CeO_2(111)$ surface, is too weak to influence the topological properties of silicene. Obviously, $\gamma_m$ is much larger than the other three parameters. Therefore, the large equivalent electric field (the "mass" term) induced by substrate is the main cause of increasing band gap of silicene.

It has been demonstrated that the system will be a topological insulator if the $\gamma_{so}$ term is dominant, whereas if $\gamma_m$ is dominant, the system is a normal insulator [56, 3]. To discuss the topological properties of silicene on $CeO_2$ (111) surface, we have calculated its edge states and $Z_2$ invariant. Fig. 5 (c) shows the band structure of zigzag nanoribbon of silicene on $CeO_2$ (111) surface (NA configuration). One can see the valence band is completely full, and there are no crossing point between valence band and conduction band. The electron in valence band needs extra energy to transfer to conduction band, suggesting that silicene on $CeO_2$ (111) surface a normal band insulator. Similarly, the transition from QSH state to band insulator induced by extra electric field has been discussed in graphene [57]. The transition can also been understand by symmetry: although the time-reversal symmetry is preserved, the substrate breaks the inversion symmetry in the perpendicular to the silicene. Moreover, the topologically trivial phase can be verified by $Z_2$ invariant. Considering the breaking of inversion symmetry in silicene/substrates, $Z_2$ invariant has been calculated by using the non-Abelian Berry connection [58], based on the evolution of the charge centers of the wannier functions (WCC), and implemented in the WannierTools code [59]. One can easily get $Z_2=0$ for NA and NB configurations (Fig. 5(d) and S6 (d)). This indicates that $CeO_2$ (111) surface induces topological phase transition of silicene from QSHE to band insulator, although the binding energy is very weak and Dirac cone in silicene is also approximately preserved.

The topological phase transition induced by substrate can also be verified by silicene put on $CaF_2$ (111) surface. The natural cleavage $CaF_2$ (111) surface has a small lattice mismatch to silicene, resulting into their six different stacking patterns with high symmetry (Fig. S7). We find that there is only weak vdWs interaction, while no strong covalent bonding interaction in silicene/$CaF_2$ (111) system, different from the silicene/$CeO_2$ (111) system in which there are two types of interfacial interaction. This can be understood by the atomic charge calculated by the Bader method, as shown in Fig. S8. In clean $CaF_2$ (111) surface, the Bader charge of F atom is about 7.77, close to 8 electrons corresponding to full occupation of the $p$ shell. On the contrary, the Bader charge of O atom at clean $CeO_2$ (111) surface is about 7.15, suggesting that the O atom has an incomplete $p$ shell structure. This means that the O atom at $CeO_2$ (111) surface is active, whereas the F atom at $CaF_2$ (111) surface is nearly unreactive. Therefore, the Si-O bond can be formed, while the Si-F bond cannot at interface. This indicates that the silicene-substrate interaction could be tuned by choosing appropriate substrates with different reactivity.

The band structures of silicene/$CaF_2$ (111) have shown in Figs. S9 and S10, in which Dirac cone in silicene is also approximately preserved. Similar with $CeO_2$ (111), the $CaF_2$ (111) surface also induces a band gap at Dirac cone of silicene, although the interaction between $CaF_2$ (111) and silicene is very weak (Table SI). Other substrate, such as monolayer BN, can also open a band gap in silicene (Fig. S11). Similarly, the opened band gap in silicene is attributed to the equivalent electric field induced by $CaF_2$ (111) surface. Thus, the silicene/$CaF_2$ (111) system is a normal insulator because $\gamma_m$ is dominant (Table SI), and $Z_2$ invariant further supports the transition from the topologically non-trivial nature to topologically trivial insulator.

The above results and others [28] have confirmed that the substrate induces the topological phase transition in silicene. In principle, the QSH state in silicene can be preserved as the substrate interaction is small enough. On the other hand, however, the interaction between silicene and substrate should be strong enough to stabilize silicene. To deal with this dilemma, the key issue is to expound the quantitative relationship between binding energy and band gap, and topological properties of silicene on a substrate. In vdWs system, the bigger the interfacial spacing, the weaker the interaction is. Fig. 6 displays the evolution of band gap and binding energy with interfacial distance of silicene/$CeO_2$ (111). Evidently, the band gap of silicene is firstly decreased rapidly with the interfacial distance ($d$) increasing from equilibrium spacing, indicating from another perspective that the interaction with substrate will result in an increase of band gap of silicene. Further increasing of the interfacial distance, its band gap decreases slowly, and then gradually increases to the value (1.6 meV) of freestanding silicene, thus creating a crossover. The evolution of $E_g$ suggests the appearance of semimetal phase ($E_g$=0) [41], when the interfacial distance reaches to a critical value ($d_c$: $d_c \in$(4.60, 4.80 Å) and (5.18, 5.38 Å) for NA and NB configurations, respectively). The semimetal phase is a critical phase, implying the transition from band insulator (d<$d_c$) to QSH state (d>$d_c$). As expected, Fig. 6 demonstrates that the binding energy between silicene and $CeO_2$ (111) surface is continuously reduced as the interfacial distance increases. As a consequence, to preserve the QSH state in silicene on $CeO_2$ (111) surface, their interfacial distance should be large enough, where their binding energy is small (about 30 meV), roughly equivalent to typical thermal energy at room temperature. This indicates that preserving both the QSH state and the stability of silicene is mutually exclusive on a substrate, due to the fact that the Dirac cone consisted of $p_z$ states is easily destroyed [39].

Since the substrate effect on the electronic structure of silicene can be equivalent to an electric field, the natural question is: whether can its QSH state be recovered by directly applying an external electric field? We calculate the band structures of silicene/CeO$_2$ (111) (NA and NB configuration) under different external electric fields, as given in Figs. S12 and S13. Disappointingly, the large band gap of silicene could hardly be tuned by external electric field, regardless of its magnitude and direction. This is due to the fact that the substrate may largely screen the extra electric field [39]. Even worse, the strong electric field will make Dirac cone in silicene into the valence or conduction band of CeO$_2$, implying the QSH state cannot be recovered by directly applying an external electric field.

**Protection of topological phase in silicene**. To facilitate its applications in spintronic device, we here first propose a practical strategy——constructing inverse symmetrical sandwich structure (protective layer /silicene/substrate) ——to maintain the topologically nontrivial nature of silicene. The CeO$_2$ (111)/silicene/CeO$_2$ (111) structures (Figs. 7 and S15 (a)) with inverse symmetry are constructed to demonstrate this method. In this kind of structure, CeO2 (111) surface is both substrate and protective layer. DFT calculation shows that their Dirac cone is very close to that of freestanding silicene (Fig. 7 (b) and S15 (b)), and their band gaps (0.6 and 3.0 meV for trilayer NA and NB configurations, respectively) are much smaller than those of silicene/CeO$_2$ (111). For trilayer structure, the extra Rashba SOC $\gamma_{R1}$, "mass" term $\gamma_m$ and "pseudomagnetic" term $\gamma_B$ are all vanished. Band structures of zigzag nanoribbon of CeO$_2$ (111)/silicene/CeO$_2$ (111) show the crossing of the edge states at the Brillouin zone boundary (Fig. 7 (c) and S15 (c)). Moreover, Z$_2$ invariant obtained by evolution of WCC (Figs. 7 (c) and S15 (d)) reveals that the CeO$_2$ (111)/silicene/CeO$_2$ (111) is in the non-trivial topological phase. The recovering of the non-trivial topology of silicene can be attributed to constructing inversion symmetry: the existence of both inversion symmetry and time-reversal symmetry ensure the topologically nontrivial nature of silicene in the sandwich structure.

The universality of this strategy can further be verified by CaF$_2$ (111)/silicene/CaF$_2$ (111), in which six configurations have been taken into account (Fig. S16). One can see from Fig. S17 that their Dirac cones with linear dispersions are clear, and their band gaps are much smaller than that of silicene/CaF$_2$ (111). Meanwhile, Z$_2$ invariant also reveals that they are in the non-trivial topological phase. More importantly, such symmetrical sandwich structure (protective layer/silicene/substrate) can not only preserve the QSH state, but also protect its stability of silicene. Recent developments in synthesis of 2D transition metal oxide nanosheets [60] and high level transfer techniques for 2D materials [61] make the integration of silicene into devices technologically feasible, including the fabrication of symmetrical sandwich structure (protective layer/silicene/substrate) proposed here. A recent experiment is also shown that the bilayer silicene was successfully obtained in a sandwich structure——$CaF_2$/bilayer silicene/$CaF_2$ [62], demonstrating that the symmetrical sandwich structure——protective layer/monolayer silicene/substrate—— to protect topological phase in silicene, can be realized experimentally.

## IV. SUMMARY

In conclusion, we have demonstrated that the magnetism in silicene can be induced by an epitaxial substrate with perfect lattice match, such as $CeO_2$ (111), and it is attributed to the breaking of the π-bonding network when half of the Si atoms are chemically bonded to the substrate. The topological phase transition in silicene occurs no matter how weak the interaction between silicene and substrate. Applying an external electric field seems impossible to recover the QSH state in silicene because of the screen effect of the substrate. We propose an inverse symmetrical sandwich structure, as an effective strategy, to preserve the QSH state, as well as the stability of silicene. This could be an important step toward development of silicene-based spintronic devices.

## ACKNOWLEDGMENTS

We thank Prof. Yugui Yao and Prof. Yabin Yu for valuable suggestions. The work was supported by the National Natural Science Foundation of China (Nos. 51772085 and 11574079).

Figures

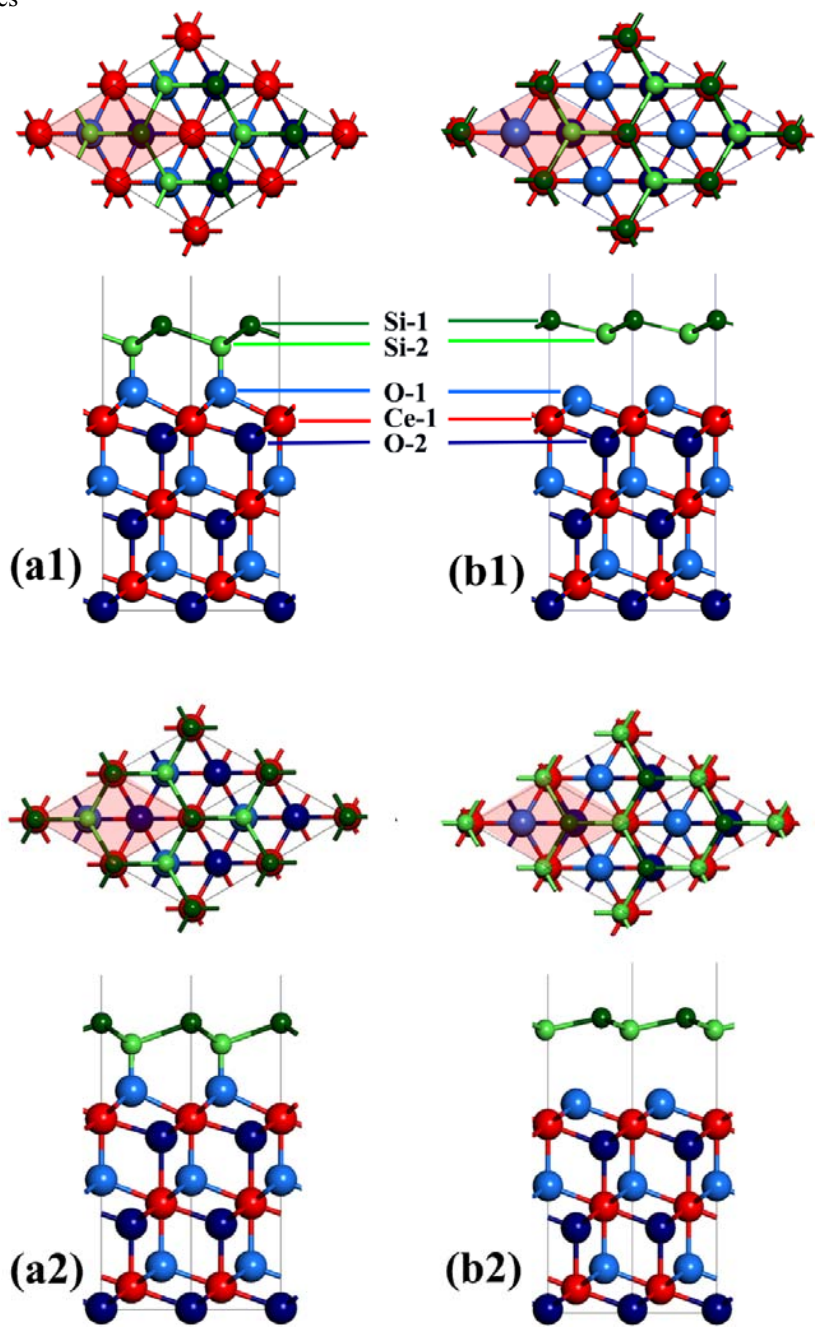

**FIG. 1.** The silicene deposited on CeO$_2$ (111) surface. (a1, a2) Covalent bonding configurations (CA, CB) and (b1, b2) non-bonding configurations (NA, NB). Red spheres represent Ce atoms, dark green and light green spheres represent top and bottom Si atoms of silicene, and dark blue and light blue spheres represent O atoms.

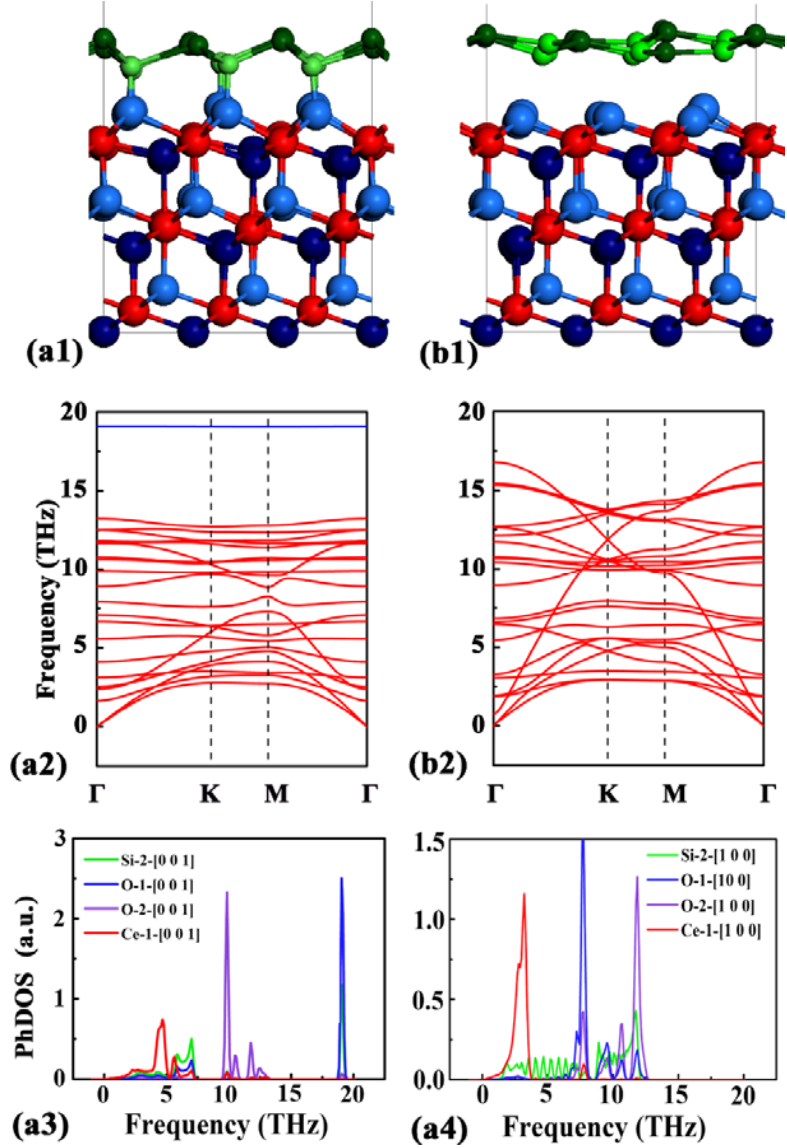

**FIG. 2.** Thermodynamic stability and bonding characters. (a1, b1) The geometrical structures of covalent bonding CA and non-bonding NA configurations, after 5 ps AIMD simulations at 300 K, suggesting the stability of silicene on $CeO_2$(111) at room. (a2, b2) Phonon dispersions of covalent bonding CA and non-bonding NA configurations. There are no image frequency in these configurations, and the high frequency mode appears in covalent bonding configuration. (a3, a4) The atom-resolved phonon density of states (PhDOS) perpendicular and parallel to the interface for CA configuration.

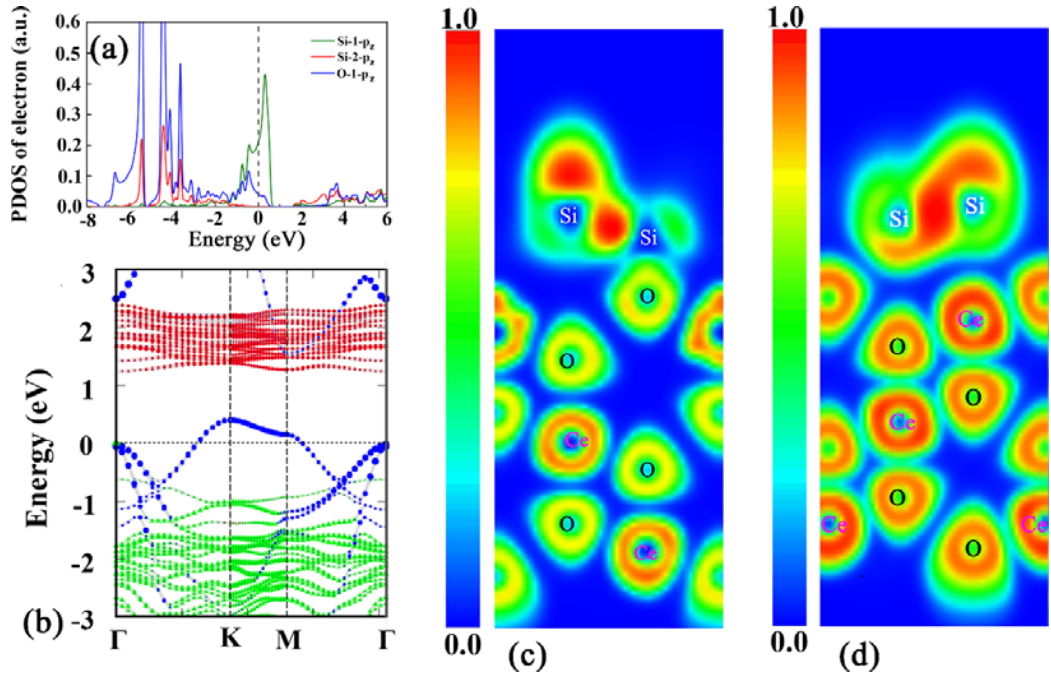

**FIG. 3.** (a) The partial density of states of electrons of covalent bonding CA configuration. (b) The projected band structure of covalent bonding CA configuration. Red, green and blue symbols represent the projected band dispersion of Ce, O and Si, respectively. (c, d) The contrasting ELFs of covalent bonding CA and non-bonding NA configurations.

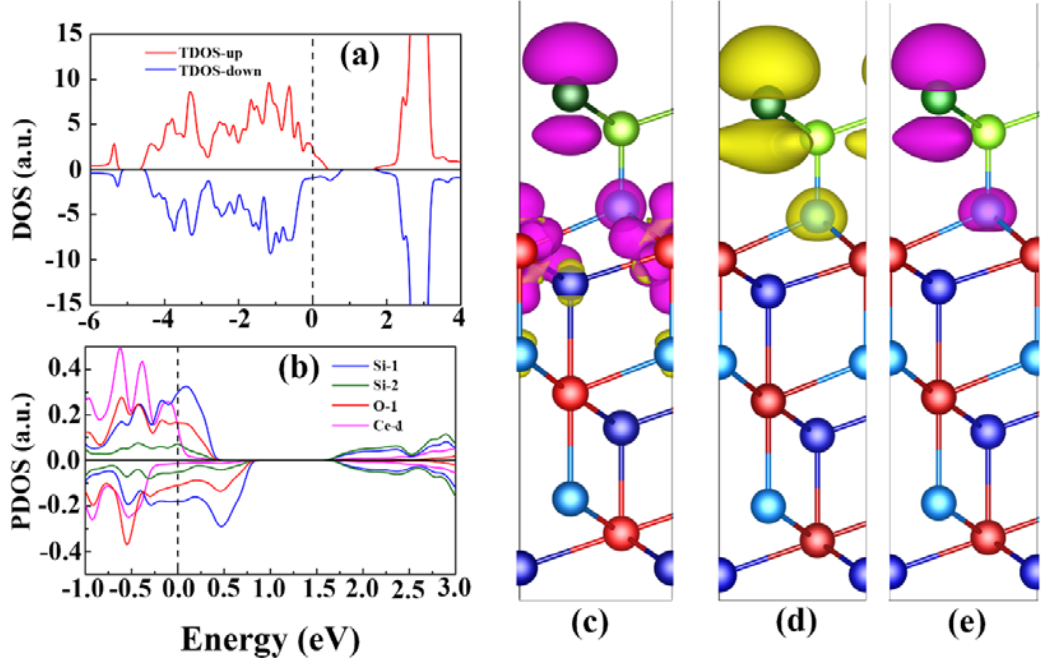

**FIG. 4.** The electronic structure of covalent bonding CA configuration. (a, b) The spin-polarized total and partial density of states of electrons. (c) Three-dimensional spatial total electron spin-density distribution $\rho_\uparrow(\vec{r}) - \rho_\downarrow(\vec{r})$. (d, e) The hole spin-density distribution at VB $\rho_\uparrow(\vec{r})$ and $\rho_\downarrow(\vec{r})$.

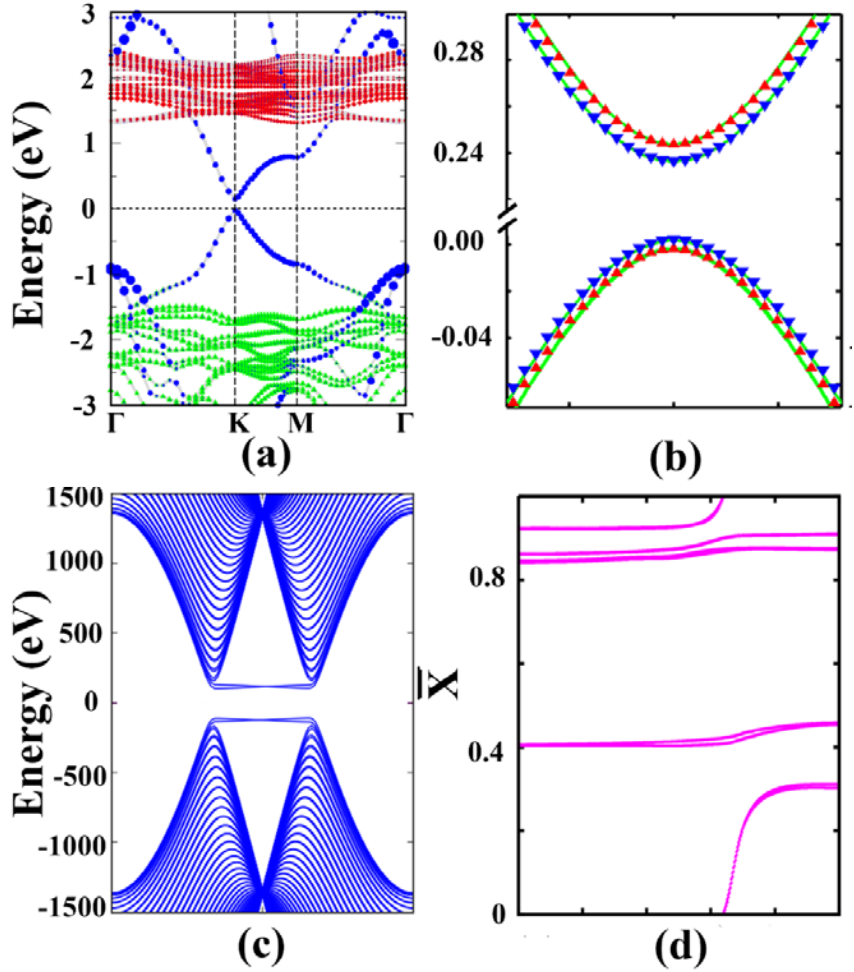

**FIG. 5.** The electronic structures and topological properties of non-bonding NA configuration. (a) Projected band structure. Red, green and blue symbols represent the projected band dispersion of Ce, O and Si, respectively. (b) Band structures near K point. Green solid lines result from tight-bind approach, triangle marks are results of first-principles calculations. (c) Band structure of zigzag nanoribbon of silicene on $CeO_2$(111). There are no edge crossing to fermi level. (d) The evolution of the charge centers of the Wannier functions of silicene on $CeO_2$(111), implying the $Z_2$ invariant is zero.

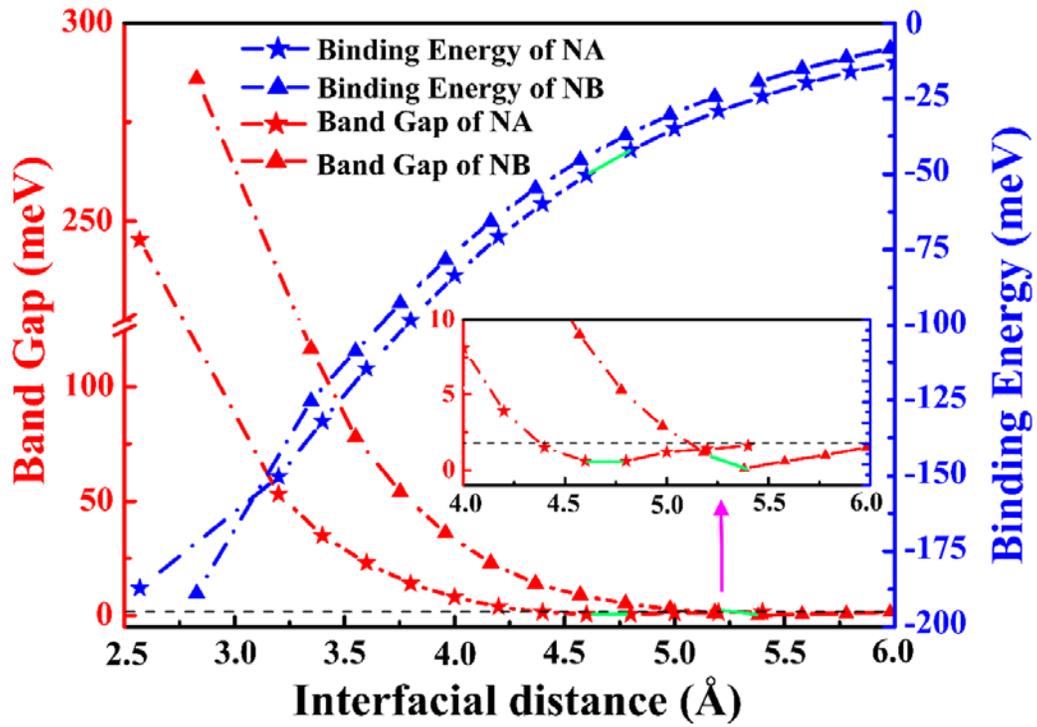

**FIG. 6.** Band gap (red line) and bind energy (blue line) as functions of interfacial distance between silicene and $CeO_2$ (111), only lower silicon in silicene atoms was constrained. Dot line represented the band insulator, while the quantum spin Hall state appear in solid line zone. Black dash line represents the band gap of freestanding silicene. The green solid line imply the topological phase transitions occur, and the semimetal state (band gap is zero) appear.

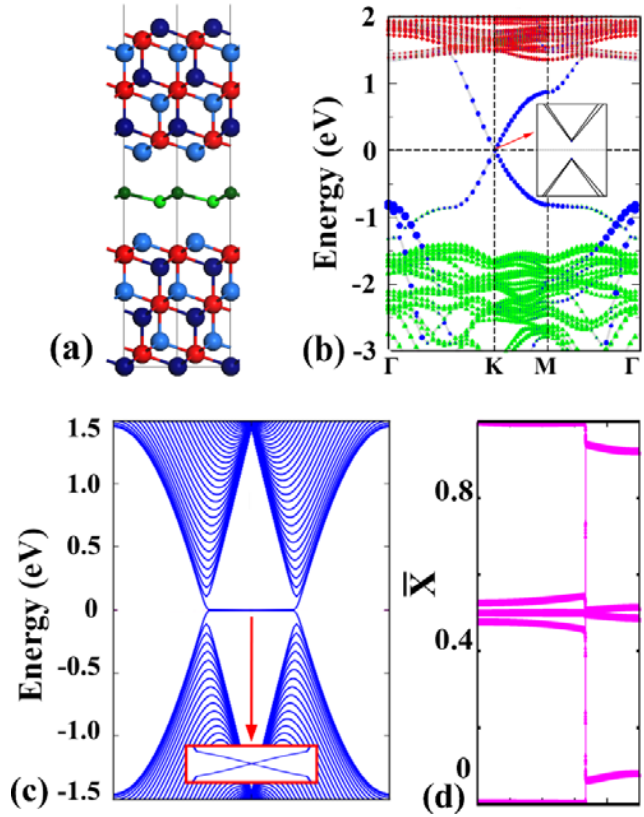

**FIG. 7.** Geometry, electronic structures and topological properties of inverse symmetrical sandwich structure corresponding to NA configuration. (a) The geometrical structure. Red spheres represent Ce atoms, dark green and light green spheres represent the upper and lower layer Si atoms of silicene, and dark blue and light blue spheres represent O atoms. (b) Band structures. Red, green and blue symbols represent the projected band dispersion of Ce, O and Si. (c) Band structure of zigzag nanoribbon of silicene in $CeO_2$ (111)/silicene/$CeO_2$ (111). (d) The evolution of the charge centers of the Wannier functions, implying the $Z_2$ invariant is one.

Table I. Silicene on the CeO$_2$ (111) surface. The E$_b$ is the binding energy per Si atom. The E$_g$ is the gap calculated at the K point. The Hamiltonian parameters defined in Eqs. (1) are given in meV. The b is the separation between the upper and lower layer silicene. The d is the interfacial distance between silicene and substrate.

|  | E$_b$ (meV) | E$_g$ (meV) | $\gamma_m$ | $\gamma_{so}$ | $\gamma_{R1}$ | $\gamma_B$ | b (Å) | $v_F$ ($10^5$m/s) | d (Å) |
|---|---|---|---|---|---|---|---|---|---|
| Silicene | / | 1.6 | / | 0.80 | / | / | 0.47 | 5.58 | / |
| NA | 374.3 | 245.2 | 119.59 | 2.78 | 10.62 | 1.16 | 0.52 | 4.55 | 2.57 |
| NA-H | 345.2 | 183.9 | 90.43 | 1.47 | 5.72 | 0.17 | 0.47 | 5.03 | 2.59 |
| NB | 377.7 | 286.1 | 142.06 | 0.95 | 4.51 | -0.36 | 0.46 | 4.86 | 2.83 |
| NB-H | 368.7 | 287.0 | 142.48 | 1.00 | 4.70 | -0.38 | 0.47 | 4.98 | 2.83 |